\documentclass[prl,showpacs,preprintnumbers,superscriptaddress,floatfix]{revtex4}
\usepackage{amsfonts}
\usepackage{graphicx}
\usepackage{graphicx,epsfig,latexsym,amssymb}
\usepackage{multirow,amsmath,array,booktabs}
\usepackage{dcolumn}
\usepackage[section]{placeins}
\usepackage{bm}

\begin{document}

\title{Exploration on the relativistic symmetry by similarity
renormalization group}
\author{Jian-You Guo}
\email[E-mail:]{jianyou@ahu.edu.cn}
\affiliation{School of Physics and Material Science, Anhui University, Hefei 230039,
People's Republic of China}

\begin{abstract}
The similarity renormalization group is used to transform Dirac Hamiltonian
into a diagonal form, which the upper(lower) diagonal element becomes an
operator describing Dirac (anti)particle. The eigenvalues of the operator
are checked in good agreement with that of the original Hamiltonian.
Furthermore, the pseudospin symmetry is investigated. It is shown that the
pseudospin splittings appearing in the non-relativistic limit are reduced by
the contributions from these terms relating the spin-orbit interactions,
added by those relating the dynamical terms, and the quality of pseudospin
symmetry origins mainly from the competition of the dynamical effects and
the spin-orbit interactions. The spin symmetry of antiparticle spectrum is
well reproduced in the present calculations.
\end{abstract}

\maketitle

Many years ago a quasidegeneracy was observed in heavy nuclei between
single-nucleon doublets with quantum numbers $(n,l,j=l+1/2)$ and $%
(n-1,l+2,j=l+3/2)$ where $n$, $l$, and $j$ are the radial, the orbital, and
the total angular momentum quantum numbers, respectively~\cite%
{Hecht69,Arima69}. The quasidegenerate states were suggested to be
pseudospin doublets $j=\tilde{l}\pm \tilde{s}$ with the pseudo orbital
angular momentum $\tilde{l}=l+1$, and the pseudospin angular momentum $%
\tilde{s}=1/2$, and have explained a number of phenomena in nuclear
structure. Because of these successes, there have been comprehensive efforts
to understand the origin of this symmetry. Until 1997, it was identified as
a relativistic symmetry \cite{Ginoc97}. Nevertheless, there is still a large
amount of attention on this symmetry. The pseudospin symmetry (PSS) of
nuclear wave functions was tested in Refs.\cite{Ginoc98,Ginoc01} with
conclusion supporting the claim in Ref.\cite{Ginoc97}. The existence of
broken PSS was checked in Refs.\cite{Gambhir98,Lalaz98}, where the
quasidegenerate pseudospin doublets were confirmed to exist near the Fermi
surface for spherical and deformed nuclei. The isospin dependence of PSS was
investigated in Ref.~\cite{Meng99}, where it is found that PSS is better for
exotic nuclei with a highly diffuse potential. PSS was shown to be
approximately conserved in medium-energy nucleon scattering from even-even
nuclei~\cite{Ginoc99PRL,Leeb00,Leeb04}. In combination with the analytic
continuation method, the resonant states were exposed to hold the PSS in
Refs.\cite{Guo053,Guo06}. In Ref.\cite{Alberto07}, the conditions which
originate the spin and pseudospin symmetries in the Dirac equation were
shown to be the same that produce equivalent energy spectra of relativistic
spin-1/2 and spin-0 particles in the presence of vector and scalar
potentials. Furthermore, the symmetries and super-symmetries of the Dirac
Hamiltonian were checked for particle moving in the spherical or
axially-deformed scalar and vector potentials \cite{Leviatan049}. More
reviews on the PSS can be found in the literature \cite{Ginoc05PR} and the
references therein. Recently, a perturbation method was adopted to
investigate the spin and pseudospin symmetries by dividing the Dirac
Hamiltonian into the part of possessing the exact (pseudo)spin symmetry and
that of breaking the symmetry~\cite{Liang11}.

Despite the large number of studies on PSS, it is still not fully understood
the origin of PSS and its breaking mechanism since there is no bound states
in the PSS limit. Hence, many efforts are devoted to compare the
contributions of different terms in the Schr\"{o}dinger-like equation for
the lower component of Dirac spinor to the pseudospin energy splitting. In
Refs.\cite{Meng98,Sugaw98,Sugaw00}, the PSS in real nuclei was shown in
connection with the competition between the pseudo-centrifugal barrier and
the pseudospin-orbital potential. In Refs.\cite{Alber012,Lisboa10}, it was
shown that the observed pseudospin splitting arises from a cancellation of
the several energy components, and the PSS in nuclei has a dynamical
character. A similar conclusion was reached in Refs.~\cite{Marco01,Marco08}.
However, in these studies, one encounters inevitably the singularity in
calculating the contribution of every component to the pseudospin splitting,
and the coupling between the energy $\epsilon $ and the operator in
solving the Schr\"{o}dinger-like equation for the lower component of Dirac
spinor (to see Eq.(\ref{lower}) in the following), which affect our
understanding on the origin of the PSS. As seen in the following Eq.(\ref%
{lower}), it seems that only $\frac{\kappa }{r}\frac{\Sigma ^{\prime }}{%
4M_{-}^{2}}$ destroys the PSS, but the pseudospin splittings are related to
every component \cite{Alber012,Lisboa10,Marco01,Marco08}. In order to cure
these defects, in the paper we transform the Dirac operator into a diagonal
form by similarity renormalization group (SRG), in which the upper(lower)
diagonal part becomes an operator describing Dirac (anti)particle with the
singularity and the coupling disappearing. In the following, we first derive
out the operator, and then present its application in analyzing the PSS.

Assuming the spherical symmetry, the radial Dirac equation can be cast in
the form of
\begin{equation}
H_{s}\psi =\epsilon \psi ,  \label{Diraceq}
\end{equation}%
with
\begin{equation}
H_{s}=\left(
\begin{array}{cc}
M+\Sigma (r) & -\frac{d}{dr}+\frac{\kappa }{r} \\
\frac{d}{dr}+\frac{\kappa }{r} & -M+\Delta (r)%
\end{array}%
\right) \text{ \ and }\psi =\left(
\begin{array}{c}
F\left( r\right) \\
G\left( r\right)%
\end{array}%
\right) ,  \label{hamilton}
\end{equation}%
where $\Sigma (r)=V(r)+S(r)$ and $\Delta (r)=V(r)-S(r)$ denote the
combinations of the scalar potential $S(r)$ and the vector potential $V(r)$,
and $\kappa $ is defined as $\kappa =(l-j)(2j+1)$. To understand the PSS,
one decouples Eq.(\ref{Diraceq}) into two equations for the upper and lower
components:
\begin{eqnarray}
\left[ -\frac{1}{2M_{+}}\left( \frac{d^{2}}{dr^{2}}+\frac{\Delta ^{\prime }}{%
2M_{+}}\frac{d}{dr}-\frac{\kappa (\kappa +1)}{r^{2}}\right) +\left( M+\Sigma
\right) -\frac{\kappa }{r}\frac{\Delta ^{\prime }}{4M_{+}^{2}}\right] F(r)
&=&\epsilon F(r),  \label{upper} \\
\left[ -\frac{1}{2M_{-}}\left( \frac{d^{2}}{dr^{2}}+\frac{\Sigma ^{\prime }}{%
2M_{-}}\frac{d}{dr}-\frac{\kappa (\kappa -1)}{r^{2}}\right) -\left( M-\Delta
\right) +\frac{\kappa }{r}\frac{\Sigma ^{\prime }}{4M_{-}^{2}}\right] G(r)
&=&\epsilon G(r),  \label{lower}
\end{eqnarray}%
here the effective masses $2M_{+}=\epsilon +M-\Delta $ and $2M_{-}=\epsilon
-M-\Sigma $. The prime denotes derivative with respect to $r$. From Eq.(\ref%
{lower}), it can be seen that the system possesses exact PSS when $\Sigma
^{\prime }=0$. Unfortunately, the condition cannot be realized in real
nuclei, many efforts are devoted to analyze the contributions of various
terms to the PSS~\cite{Alber012,Lisboa10,Marco01,Marco08}. However, as there
exist deficiencies mentioned before, we decouple Eq.(\ref{Diraceq}) by SRG.

Without loss of generality, we begin our formalism for a general Dirac
Hamiltonian $H=\vec{\alpha}\cdot \vec{p}+\beta (M+S)+V$, which is fully
applicable for $H_{s}$. Following Wegner's formulation of the SRG\cite%
{Wegner94}, the initial Hamiltonian $H$ is transformed by the unitary
operator $U\left( l\right) $ according to
\begin{equation}
H\left( l\right) =U\left( l\right) HU^{\dagger }\left( l\right) ,\text{ \ }%
H(0)=H  \label{unitary}
\end{equation}%
where $l$ is a flow parameter. Differentiation Eq.(\ref{unitary}) gives the
flow equation as
\begin{equation}
\frac{d}{dl}H\left( l\right) =\left[ \eta \left( l\right) ,H\left( l\right) %
\right] ,  \label{floweq}
\end{equation}%
with the generator $\eta (l)=$ $\frac{dU\left( l\right) }{dl}U^{\dagger
}\left( l\right) $. There are several possibilities to choose the $\eta (l)$
so that $H\left( l\right) $ becomes diagonal in the limit $l\rightarrow
\infty $. In Wegner's original formulation \cite{Wegner94}, $\eta (l)$ was
chosen as the commutator of the diagonal part of $H(l)$ with $H(l)$ itself,
i.e. $\eta (l)=[H_{\text{diag}}(l),H(l)]$. An alternative to Wegner's
formulation is $\eta (l)=$ $\left[ G,H(l)\right] $, where $G$ is a fixed ($l$%
-independent) hermitian operator. It is straightforward to show that $H(l)$
converges to a final Hamiltonian which commutes with $G$. Here, we hope to
transform Dirac Hamiltonian into a diagonal form, which must commute with
the $\beta $ matrix.\ Thus, it is appropriate to choose $\eta (l)$ in the
form
\begin{equation}
\eta (l)=\left[ \beta M,H(l)\right] .  \label{generator}
\end{equation}%
In order to solve Eq.(\ref{floweq}), the technique in Ref.\cite{Bylev98} is
adopted. The Hamiltonian $H(l)$ is presented as a sum of an even operator $%
\varepsilon (l)$ and odd operator $o(l)$:
\begin{equation}
\ \ \ \ H(l)=\varepsilon (l)+o(l),  \label{hamiltonian}
\end{equation}%
where the even or oddness is defined by the commutation relations of the
respective operators, i.e., $\varepsilon (l)\beta =\beta \varepsilon (l)\ $%
and $o(l)\beta =-\beta o(l)$. To put Eqs.(\ref{generator}) and (\ref%
{hamiltonian}) into Eq.(\ref{floweq}) gives
\begin{eqnarray}
\frac{d\varepsilon (l)}{dl} &=&4M\beta o^{2}(l),  \label{even} \\
\frac{do(l)}{dl} &=&2M\beta \left[ o(l),\varepsilon (l)\right] .  \label{odd}
\end{eqnarray}%
The system of Eqs.(\ref{even}) and (\ref{odd}) can be solved perturbatively
in $1/M$ \cite{Bylev98}. It is convenient to introduce a dimensionless flow
parameter $\lambda =lM^{2}$. Since $\varepsilon (0)=\beta \left( M+S\right)
+V$, and $o(0)=\vec{\alpha}\cdot \vec{p}$ and the expansion of $\varepsilon
(\lambda )/M$ in a series in $1/M$ contains terms starting with the zeroth
order term
\begin{equation}
\frac{1}{M}\varepsilon \left( \lambda \right) =\sum\limits_{i=0}^{\infty }%
\frac{1}{M^{i}}\varepsilon _{i}\left( \lambda \right) ,  \label{Expandeven}
\end{equation}%
whereas the expansion of $o(\lambda )/M$ starts with the first order
\begin{equation}
\frac{1}{M}o\left( \lambda \right) =\sum\limits_{j=1}^{\infty }\frac{1}{M^{j}%
}o_{j}\left( \lambda \right) .  \label{Expandodd}
\end{equation}%
Differentiation Eqs.(\ref{Expandeven}) and (\ref{Expandodd}) yield the
following equations,
\begin{eqnarray}
\frac{d\varepsilon _{n}\left( \lambda \right) }{d\lambda } &=&4\beta
\sum\limits_{k=1}^{n-1}o_{k}\left( \lambda \right) o_{n-k}\left( \lambda
\right) ,  \label{difeven} \\
\frac{do_{n}\left( \lambda \right) }{d\lambda } &=&-4o_{n}\left( \lambda
\right) +2\beta \sum\limits_{k=1}^{n-1}\left[ o_{k}\left( \lambda \right)
,\varepsilon _{n-k}\left( \lambda \right) \right] .  \label{difodd}
\end{eqnarray}%
The solutions of the equations (\ref{difeven}) and (\ref{difodd}) are
obtained as
\begin{eqnarray}
\varepsilon _{n}\left( \lambda \right) &=&\varepsilon _{n}\left( 0\right)
+4\beta \int\limits_{0}^{\lambda }d\lambda ^{\prime
}\sum\limits_{k=1}^{n-1}o_{k}\left( \lambda ^{\prime }\right) o_{n-k}\left(
\lambda ^{\prime }\right) ,  \label{solueven} \\
o_{n}\left( \lambda \right) &=&o_{n}\left( 0\right) e^{-4\lambda }+2\beta
e^{-4\lambda }\int\limits_{0}^{\lambda }d\lambda ^{\prime
}\sum\limits_{k=1}^{n-1}\left[ e^{4\lambda ^{\prime }}o_{k}\left( \lambda
^{\prime }\right) ,\varepsilon _{n-k}\left( \lambda ^{\prime }\right) \right]
\label{soluodd}
\end{eqnarray}%
with the initial conditions
\begin{eqnarray}
\varepsilon _{0}(0) &=&\beta ,\varepsilon _{1}(0)=\beta S+V,\varepsilon
_{n}(0)=0\text{ \ \ if \ \ }n\geqslant 2,  \notag \\
o_{1}(0) &=&\vec{\alpha}\cdot \vec{p},o_{n}(0)=0\text{ \ \ if \ \ }%
n\geqslant 2.  \label{inicondi}
\end{eqnarray}%
From the equations (\ref{solueven}-\ref{soluodd}) with the initial condition
(\ref{inicondi}), we obtain $\varepsilon _{0}(\lambda )=\beta $, $%
\varepsilon _{1}\left( \lambda \right) =\beta S+V$, and $o_{1}\left( \lambda
\right) =o_{1}\left( 0\right) e^{-4\lambda }$. Hence, it is easy to verify
that $o_{n}(\lambda )$ exponentially goes to zero when $\lambda \rightarrow
\infty $. So, the diagonalized Dirac operator is obtained as
\begin{eqnarray*}
\varepsilon (\infty ) &=&M\varepsilon _{0}(\infty )+\varepsilon _{1}(\infty
)+\frac{1}{M}\varepsilon _{2}(\infty )+\frac{1}{M^{2}}\varepsilon
_{3}(\infty )+\frac{1}{M^{3}}\varepsilon _{4}(\infty )+\cdots \\
&=&M\varepsilon _{0}(0)+\varepsilon _{1}\left( 0\right) +\frac{1}{2M}\beta
o_{1}^{2}(0)+\frac{1}{8M^{2}}\left[ \left[ o_{1}\left( 0\right) ,\varepsilon
_{1}\left( 0\right) \right] ,o_{1}\left( 0\right) \right] \\
&&+\frac{1}{32M^{3}}\beta \left( -4o_{1}^{4}\left( 0\right) +o_{1}\left(
0\right) \left[ \left[ o_{1}\left( 0\right) ,\varepsilon _{1}\left( 0\right) %
\right] ,\varepsilon _{1}\left( 0\right) \right] +\left[ \left[ o_{1}\left(
0\right) ,\varepsilon _{1}\left( 0\right) \right] ,\varepsilon _{1}\left(
0\right) \right] o_{1}\left( 0\right) -2\left[ o_{1}\left( 0\right)
,\varepsilon _{1}\left( 0\right) \right] \left[ o_{1}\left( 0\right)
,\varepsilon _{1}\left( 0\right) \right] \right) +\cdots
\end{eqnarray*}%
Here, only a spherical system is considered, $\varepsilon _{1}(0)=\left(
\begin{array}{cc}
\Sigma \left( r\right) & 0 \\
0 & \Delta \left( r\right)%
\end{array}%
\right) ,$ $o_{1}(0)=\left(
\begin{array}{cc}
0 & -\frac{d}{dr}+\frac{\kappa }{r} \\
\frac{d}{dr}+\frac{\kappa }{r} & 0%
\end{array}%
\right) $, the diagonalized Dirac operator becomes
\begin{equation}
\varepsilon (\infty )=\left(
\begin{array}{cc}
H_{1}+M & 0 \\
0 & H_{2}-M%
\end{array}%
\right) ,  \label{diagDirac}
\end{equation}%
where
\begin{eqnarray}
H_{1} &=&\Sigma \left( r\right) +\frac{p^{2}}{2M}-\frac{1}{2M^{2}}\left(
Sp^{2}-S^{\prime }\frac{d}{dr}\right) -\frac{\kappa }{r}\frac{\Delta
^{\prime }}{4M^{2}}+\frac{\Sigma ^{\prime \prime }}{8M^{2}}  \notag \\
&&+\frac{S}{2M^{3}}\left( Sp^{2}-2S^{\prime }\frac{d}{dr}\right) +\frac{%
\kappa }{r}\frac{S\Delta ^{\prime }}{2M^{3}}-\frac{{\Sigma ^{\prime }}%
^{2}-2\Sigma ^{\prime }\Delta ^{\prime }+4S\Sigma ^{\prime \prime }}{16M^{3}}%
-\frac{p^{4}}{8M^{3}}  \label{Diracp}
\end{eqnarray}%
is an operator describing Dirac particle with $p^{2}=-\frac{d^{2}}{dr^{2}}+%
\frac{\kappa \left( \kappa +1\right) }{r^{2}}$, and%
\begin{eqnarray}
H_{2} &=&\Delta \left( r\right) -\frac{p^{2}}{2M}+\frac{1}{2M^{2}}\left(
Sp^{2}-S^{\prime }\frac{d}{dr}\right) +\frac{\kappa }{r}\frac{\Sigma
^{\prime }}{4M^{2}}+\frac{\Delta ^{\prime \prime }}{8M^{2}}  \notag \\
&&-\frac{S}{2M^{3}}\left( Sp^{2}-2S^{\prime }\frac{d}{dr}\right) -\frac{%
\kappa }{r}\frac{S\Sigma ^{\prime }}{2M^{3}}+\frac{{\Delta ^{\prime }}%
^{2}-2\Sigma ^{\prime }\Delta ^{\prime }-4S\Delta ^{\prime \prime }}{16M^{3}}%
+\frac{p^{4}}{8M^{3}}  \label{ADiracp}
\end{eqnarray}%
is an operator describing Dirac antiparticle with $p^{2}=-\frac{d^{2}}{dr^{2}%
}+\frac{\kappa \left( \kappa -1\right) }{r^{2}}$. The Hamiltonian for Dirac
antiparticle in this case is in fact $-H_{2}$ with eigenvalue $-\epsilon $.
This is consistent with the transformations of potentials $S\rightarrow S$, $%
\Sigma \rightarrow -\Delta $, and $\Delta \rightarrow -\Sigma $ under charge
conjugation from $H_{1}$\cite{Lisboa10}. Here, the primes have the same
meaning as that in Eq.(\ref{lower}) and the double primes denote
second-order derivatives with respect to $r$.

The first two terms of $H_1(H_2)$ correspond to the operator describing
Dirac (anti)particle in the non-relativistic limit. The relativistic effect
begins to show up from the order of $1/M^{2}$ in the perturbation expansion
of $\varepsilon (\infty)$, which are presented in $H_1(H_2)$ from the third
to fifth terms. In order to obtain better result, the perturbation expansion
up to order $1/M^{3}$ is also included in Eqs.(\ref{Diracp}) and (\ref%
{ADiracp}).

In Eq.(\ref{Diracp}), it can be seen the spin symmetry is exact for Dirac
particle when $\Delta ^{\prime }=0$, which agrees with Eq.(\ref{upper}). The
same result is obtained for Dirac antiparticle in comparing with Ref.\cite%
{Lisboa10,Zhou03}, which can be observed from Eq.(\ref{ADiracp}) with $%
\Sigma^{\prime }=0$. Particularly, the singularity disappears in every
component of Eqs.(\ref{Diracp}) and (\ref{ADiracp}), the operators $H_{1}$
and $H_{2}$ are Hermitian. In addition, there is no the coupling between the
energy $\epsilon $ and the operator $H_{1}$($H_{2}$). Thus, the energy
spectra of $H_{1}$ and $H_{2}$ can be calculated conveniently.

The energy spectra of $H_{1}$($H_{2}$) agree the results of Eq.(\ref{Diraceq}%
) very well, and the energy splittings of pseudospin partners are in
agreement with the exact relativistic case. Especially, the contribution of
every component to the pseudospin splittings can be calculated, which is
helpful to analyze the origin of PSS. In order to convince the conclusion, a
Woods-Saxon type potential is adopted for $\Sigma (r)$ and $\Delta (r)$,
i.e., $\Sigma (r)=\Sigma _{0}f(a_{\Sigma },r_{\Sigma },r)$ and $\Delta
(r)=\Delta _{0}f(a_{\Delta },r_{\Delta },r)$ with
\begin{equation}
f(a_{0},r_{0},r)=\frac{1}{1+\exp \left( \frac{r-r_{0}}{a_{0}}\right) }\text{.%
}
\end{equation}%
The corresponding parameters are determined by fitting the energy spectrum
from the RMF calculations for $^{208}$Pb (to see Ref.~\cite{Guo053}). The
energy spectra of $H_{1}$($H_{2}$) are calculated by expansion in harmonic
oscillator basis. The energy spectra of the six pseudospin partners are
shown in Fig.1, where the first column in each subfigure corresponds to that
$H_{1}$ is approximated to the non-relativistic limit. The second and third
columns in each subfigure correspond to that $H_{1}$ is approximated to the
order $1/M^{2}$ and $1/M^{3}$, respectively. The exact relativistic spectra
(the eigenvalues of Eq.(\ref{Diraceq})) are displayed in the fourth column.
From Fig.1, it can be seen that the deviations between the non-relativistic
limit (the first column) and the exact relativistic case (the fourth column)
are very large, i.e., the relativistic effect is apparent in the present
system. With the increasing perturbation order, the calculated result is
closer to the exact relativistic one. When $H_{1}$ is approximated to the
order $1/M^{3}$, the calculated spectra are considerably agreeable with
those from the exact relativistic calculations. Especially, the pseudospin
energy splitting is in good agreement with the exact relativistic result.
These show the operator $H_{1}$ presents a good description for Dirac
particle, and can be used to analyze the PSS.

In order to disclose the origin of PSS, we check the contribution of every
component in $H_{1}$ to the pseudospin energy splitting. Based on the
consideration of attribute and hermitian, we decompose $H_{1}$ into the
eight components: $\Sigma \left( r\right) +\frac{p^{2}}{2M}$, $-\frac{1}{%
2M^{2}}\left( Sp^{2}-S^{\prime }\frac{d}{dr}\right) $, $-\frac{\kappa }{r}%
\frac{\Delta ^{\prime }}{4M^{2}}$, $\frac{\Sigma ^{\prime \prime }}{8M^{2}}$%
, $\frac{S}{2M^{3}}\left( Sp^{2}-2S^{\prime }\frac{d}{dr}\right) $, $\frac{%
\kappa }{r}\frac{S\Delta ^{\prime }}{2M^{3}}$, $-\frac{{\Sigma ^{\prime }}%
^{2}-2\Sigma ^{\prime }\Delta ^{\prime }+4S\Sigma ^{\prime \prime }}{16M^{3}}
$, $-\frac{p^{4}}{8M^{3}}$, which are respectively labelled as $%
O_{1},O_{2},\cdots ,O_{8}$. For the $k$-state with eigenvector $\psi _{k}$,
the contribution of $O_{i}$ to the level $E_{k}$ is calculated by the
formula $\left\langle k\right\vert O_{i}\left\vert k\right\rangle
=\int_{0}^{\infty }\psi _{k}^{\ast }O_{i}\psi _{k}d^{3}\vec{r}$, which is
denoted as $\epsilon _{i}(k)$. To reduce the length of the article, only the
data for two pseudospin doublets are listed in Table I.

The contribution of $O_{1}$ to $\Delta \epsilon $ is very large, which means
that the spectra of $H_{1}$ in the non-relativistic limit do not have the
PSS. Namely, PSS is not a non-relativistic symmetry. It has a relativistic
origin, which agrees with the claim in Ref.\cite{Ginoc97}. The contributions
of $O_{3}$ and $O_{6}$ to $\Delta \epsilon $ are negative, which implies the
pseudospin splittings in the non-relativistic limit are reduced by the
contributions from these terms relating the spin-orbit interactions, and
agrees with the relativistic interpretation of PSS. The contributions of $%
O_{2}$ and $O_{5}$ to $\Delta \epsilon $ are positive, which means the
pseudospin splittings are added by the contributions relating the dynamical
terms, and supports a relativistic origin for this symmetry. Compared with $%
O_{2}$, $O_{3}$, $O_{5}$ and $O_{6}$, the contributions of $O_{4}$, $O_{7}$
and $O_{8}$ to $\Delta \epsilon $ are relatively minor. These show the
quality of PSS origins mainly from the competition of the spin-orbit
interactions and the dynamical effects. Although the contributions of $O_{4}$%
, $O_{7}$ and $O_{8}$ to $\Delta \epsilon $ are minor, their influences on
PSS can not be ignored, which supports partly with the claim in Refs.\cite%
{Alber012,Lisboa10,Marco01}, i.e., the observed pseudospin splitting arises
from a cancellation of the several energy components, and the PSS in nuclei
has a dynamical character.

In order to check further the applicability and validity for the present
formalism, we have calculated the energy spectrum of $H_{2}$ for Dirac
antiparticle, which is shown in Fig.2. The good spin symmetry is displayed
clearly, which is in agreement with Refs.~\cite{Lisboa10,Zhou03}.

In summary, the similarity renormalization group is used to transform the
spherical Dirac operator into a diagonal form. The upper(lower) diagonal
element becomes an operator describing Dirac (anti)particle, which holds the
form of Schr\"{o}dinger-like operator with the singularity disappearing in
every component. The energy spectra of the operator are calculated in good
agreement with the exact relativistic ones. By comparing the contributions
of the various components to the energy splittings, PSS is shown to be a
relativistic symmetry. The quality of PSS is correlated with the
contribution of every component of $H_{1}$ to the pseudospin splitting,
especially, the competition of the spin-orbit interactions and the dynamical
effects, which supports the claim of a dynamical character. The spin
symmetry of antiparticle spectrum is also well reproduced in the present
calculations.

Helpful discussions with professor Zhou and Doctor Lv are acknowledged. We
would like to thank the anonymous referee for pointing out a sign error.
This work was partly supported by the National Natural Science Foundation of
China under Grant No.11175001, the Excellent Talents Cultivation Foundation
of Anhui Province under Grant No.2007Z018, the Natural Science Foundation of
Anhui Province under Grant No.11040606M07, and the 211 Project of Anhui
University.


\begin{table}[tbp]
\caption{{The contribution of the operator $O_i$ to the level $E_k$: $%
\protect\epsilon _{i}(k)= \left\langle k\right\vert O_{i}\left\vert
k\right\rangle$, where $O_i$ can be seen in text, $k$ represents the single
particle states $2d_{5/2}$, $1g_{7/2}$, $2f_{7/2}$, and $1h_{9/2}$, and $%
\Delta\protect\epsilon=\protect\epsilon_i(a)-\protect\epsilon_i(b)$. The
data listed in the last line is a sum from the first line to the eighth
line. }}%
\begin{tabular}{c|ccc|ccc}
\hline\hline
& $2d_{5/2}(a)$ & $1g_{7/2}(b)$ &  & $2f_{7/2}(a)$ & $1h_{9/2}(b)$ &  \\
\hline
$i$ & $\epsilon_i(a)$ & $\epsilon_i(b)$ & $\Delta\epsilon$ & $\epsilon_i(a)$
& $\epsilon_i(b)$ & $\Delta\epsilon$ \\
1 & -30.872 & -35.649 & 4.777 & -21.303 & -27.035 & 5.732 \\
2 & 7.511 & 6.746 & 0.765 & 8.448 & 8.003 & 0.445 \\
3 & -0.509 & 0.721 & -1.230 & -0.732 & 1.083 & -1.815 \\
4 & 0.018 & 0.042 & -0.023 & 0.004 & 0.042 & -0.038 \\
5 & 2.732 & 2.349 & 0.383 & 3.021 & 2.717 & 0.303 \\
6 & -0.240 & 0.410 & -0.650 & -0.303 & 0.581 & -0.884 \\
7 & 0.001 & 0.018 & -0.017 & -0.006 & 0.015 & -0.022 \\
8 & -0.316 & -0.257 & -0.059 & -0.440 & -0.389 & -0.051 \\
total & -21.675 & -25.621 & 3.946 & -11.312 & -14.982 & 3.671 \\ \hline\hline
\end{tabular}%
\end{table}

\begin{figure}[tbp]
\includegraphics[width=8cm]{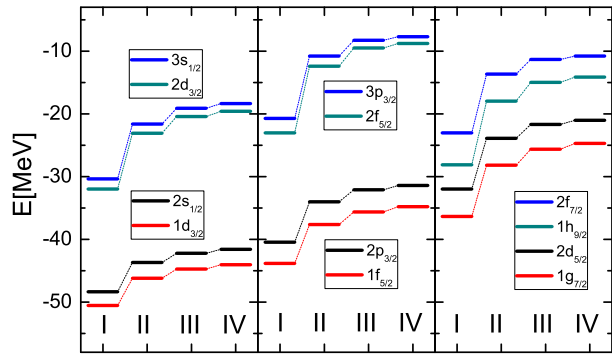}
\caption{(Color online) The energy spectrum of $H_{1}$ for the six
pseudospin partners. The first column in each subfigure corresponds to that $%
H_{1}$ is approximated to the non-relativistic limit. The second and third
columns in each subfigure correspond to that $H_{1}$ is approximated to the
order $1/M^{2}$ and $1/M^{3}$, respectively. }
\end{figure}

\begin{figure}[tbp]
\includegraphics[width=7cm]{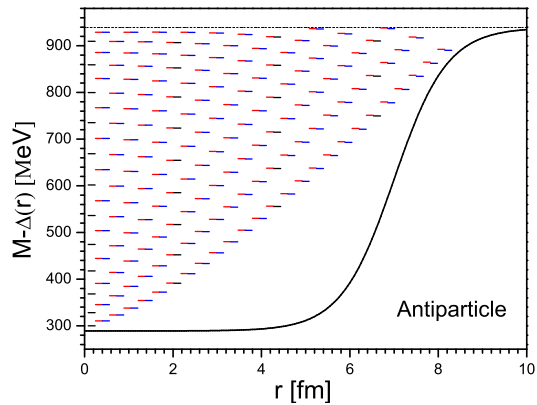}
\caption{(Color online) The energy spectrum of $H_2$ for Dirac antiparticle.}
\end{figure}

\end{document}